\documentclass[a4paper,twocolumn,superscriptaddress,floatfix,aps]{revtex4}

\usepackage[latin1]{inputenc}
\usepackage{amssymb,amsmath,amsfonts}
\usepackage{array,dcolumn,bm}
\usepackage{graphicx,xcolor}

\begin{document}

\title{Electronic cooling of a submicron-sized metallic beam}

\author{J. T. Muhonen}
\email{juha.muhonen@tkk.fi}
\affiliation{Low Temperature Laboratory, Helsinki University of Technology, P.O. BOX 3500, 02015 TKK, Finland}

\author{A. O. Niskanen}
\affiliation{VTT Technical Research Centre of Finland, Sensors, P.O. BOX 1000, 02044 VTT, Finland}

\author{M. Meschke}
\affiliation{Low Temperature Laboratory, Helsinki University of Technology, P.O. BOX 3500, 02015 TKK, Finland}

\author{Yu. A. Pashkin}
\altaffiliation{On leave from Lebedev Physical Institute, Moscow 119991, Russia}
\affiliation{NEC Nano Electronics Research Laboratories and RIKEN Advance Science Institute,
34 Miyukigaoka, Tsukuba, Ibaraki 305-8501, Japan.}

\author{J. S. Tsai}
\affiliation{NEC Nano Electronics Research Laboratories and RIKEN Advance Science Institute,
34 Miyukigaoka, Tsukuba, Ibaraki 305-8501, Japan.}

\author{L. Sainiemi}
\affiliation{Micro and Nanosciences Laboratory, Helsinki University of Technology, P.O. BOX 3500, 02015 TKK, Finland}

\author{S. Franssila}
\affiliation{Micro and Nanosciences Laboratory, Helsinki University of Technology, P.O. BOX 3500, 02015 TKK, Finland}

\author{J. P. Pekola}
\affiliation{Low Temperature Laboratory, Helsinki University of Technology, P.O. BOX 3500, 02015 TKK, Finland}

\begin{abstract}
We demonstrate electronic cooling of a suspended AuPd island using superconductor-insulator-normal metal tunnel junctions.
This was achieved by developing a simple fabrication method for reliably releasing narrow submicron sized metal beams.
The process is based on reactive ion etching and uses a conducting substrate to avoid charge-up damage and is compatible with e.g. conventional e-beam lithography, shadow-angle metal deposition and oxide tunnel junctions.
The devices function well and exhibit clear cooling; up to factor of two at sub-kelvin temperatures.
%We observe no notable effect from the reduced dimensionality of the normal metal island.
\end{abstract}

\date{\today}

\maketitle

Nanomechanical devices \cite{cleland03,blencowe} at low temperatures are currently a topic of intense study \cite{schwab05}. One motivation for this is the possibility to drive these devices into their quantum limit by cooling down their relevant mechanical modes below the quantum of resonator energy \cite{lahaye04}. To actually perform nontrivial quantum mechanical experiments it is then necessary to couple the resonator to some other elements. A large class of interesting objects are metallic nanoelectronic devices, either superconducting or normal metal. Many groups have approached the task at hand by separately fabricating a non-conducting, possibly single crystal, mechanical device and then metallizing it \cite{roukes03,naik06,roukes08}. This has been done also with SINIS structures \cite{Yung03}. Surprisingly, evaporated metal beams can have almost as good mechanical properties \cite{LiNEMS,regal} as their more obvious alternatives. Integrating a metallic beam with a solid state refrigerator is therefore an intriguing prospect. Suspended micrometer scale metallic beams are in use in the field of bolometers \cite{luukanen} and in microelectromechanical systems (MEMS) \cite{karim07, song07}. However, only a few examples \cite{Sorin, LiSET} of suspended metallic nanostructures with tunnel junctions have been fabricated so far.

\begin{figure}
\begin{center}
\includegraphics[width=1\linewidth]{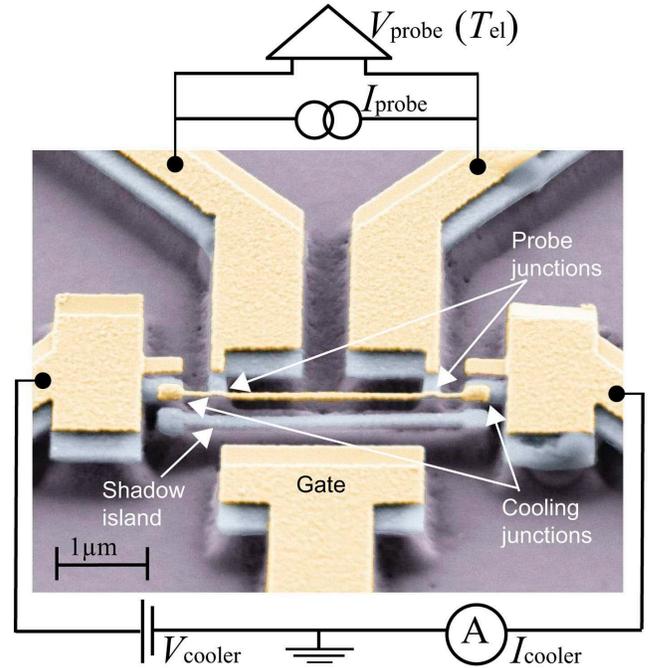} 
\caption{Suspended SINIS cooler and measurement schematics. AuPd (light) island is connected to Al (gray) leads through tunnel junctions. The larger junctions are voltage biased for cooling and the smaller ones current biased for thermometry. The gate electrode in the bottom was not used in the present measurements.}
\label{fig1}
\end{center}
\end{figure}

With the motivation of studying the effect of suspension on the properties of normal metal-insulator-superconductor (NIS) electron coolers \cite{nahum,leivo}, we have developed a reliable and straightforward fabrication process for fully metallic suspended structures. The process is also useful for fabricating other kinds of partly suspended metallic circuits which require tunnel junctions, e.g. qubit circuits. Our fabrication process begins by e-beam patterning and angle evaporating a circuit on a conducting substrate, in our case n-doped (3-5 $\Omega$cm) unoxidized silicon. The normal metal used is an alloy of Au (3 parts) and Pd (1 part) and the superconducting material is Al. Al is deposited first and then oxidized to form the tunnel barriers. After this, AuPd is deposited from a different angle to form the island. Film thicknesses are 65 nm for Al and 50 nm for AuPd and the island width is about 90 nm. After lift-off, we release the narrow parts of the pattern from the substrate by isotropic etching using SF$_6$ in a reactive ion etcher (RIE). Pressure of 100 mTorr, power of 80 W, SF$_6$ flow of 30 sccm and O$_2$ flow of 5 sccm were used. These parameters produced a nicely isotropic etching profile which allows us to selectively suspend only the narrowest parts by using suitable etching time, which was roughly 1 minute in our case. With this method we fabricated SINIS refrigerators in which only the normal metal island is released from the substrate as depicted in Fig.~\ref{fig1}. Naturally, the edges of wide structures are also suspended.

High-resistivity Si, and Si with a 300 nm thick SiO$_2$ layer on top of it, were first tried as substrate materials but it turned out that with these insulating materials always at least one of the junctions was shorted after etching. This might be due to the so-called antenna effect that has been studied before \cite{charge1, charge2} in the context of gate oxide damage in semiconductors. Metal films directly in contact with the plasma can act as charge collecting antennas during etching, hence, causing voltage to build up over oxides separating them from other layers. If we consider that in our samples the bonding pads are acting as antennas and are connected to the Al leads, assuming charge collection of 10 pA/$\mu$m$^2$ \cite{charge2} could lead to a voltage variation of some volts over the junctions. Using conducting substrate, in our case n-doped Si, suppresses this effect as the junctions are effectively shunted through the substrate.

The theoretical cooling power and electric current of SINIS structures are quite well understood, see e.g. Ref.~\cite{giazotto06} for a review and formulas. Qualitatively, no current flows if the voltage over each NIS junction is below $\Delta/e$, i.e., $2\Delta/e$ for two junctions in series, where $\Delta$ is the superconducting gap. When a voltage bias is applied across the junction so that the Fermi level of the normal metal part is shifted upwards by energy $eV\approx\Delta$, current can start flowing. If the bias voltage is just below the gap, the tunnel junctions serve as energy filters by selectively allowing, depending on polarity, either hot electrons to exit or cool ones to enter the normal metal island. Since there are two such junctions in series, both these processes coexist. This then leads to sharpening of the Fermi distribution of the electron gas i.e. cooling. When the bias voltage is much above the gap value, the junction will start to act as a normal resistor with linear current-voltage (I-V) behavior and Joule heating appears. At elevated temperatures the I-V characteristics get increasingly rounded at the gap edges. This effect is used for thermometry by current biasing the smaller junctions with a small probe current ($\sim$pA) and measuring the induced voltage, which approaches $2\Delta/e$ when $T \ll \Delta/k_B$ and zero at high temperatures. Figure~\ref{fig1} illustrates the basic measurement schematics. The resistances of the small junctions are approximately 65 k$\Omega$ and the larger cooling junctions have resistance of about 10 k$\Omega$.

\begin{figure}
\begin{center}
\includegraphics[width=1\linewidth]{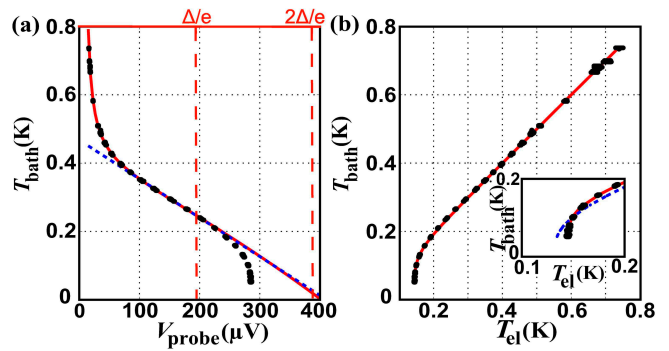} 
\caption{(\textbf{a}) Calibration of the integrated electron thermometer. The measured data ($V_{\rm cooler}=0$) is shown by dots. The dashed line is a linear fit to the center region. The solid line is the calibration used in this paper. Note that the exact normal metal temperature values below 100 mK or so may not be fully accurate (in (b) and Fig.~\ref{fig3}) due to extrapolation. The asymptotic linear behavior and intersection at about $2\Delta/e$ are quantitatively correct though. The thermometer loses sensitivity at high temperatures.
(\textbf{b}) Dependence of island temperature $T_{\rm el}$ on the bath temperature $T_{\rm bath}$ at zero cooler bias. The normal metal island does not cool below 145 mK even when the cryostat (phonon) temperature is reduced below this. The solid line is a fit assuming conventional bulk electron-phonon coupling. The inset is a close-up on the low temperature end showing a fit with the $T^3$ electron-phonon coupling as a dashed line.}
\label{fig2}
\end{center}
\end{figure}

To calibrate the thermometer, we sweep the bath temperature of our dilution refrigerator while passing approximately a 12 pA probe current from a dc current source through the probe junctions. We then record the probe voltage $V_{\rm probe}$ at different bath temperatures, this is shown in Fig.~\ref{fig2}(a). The cooler junctions are biased with a zero voltage such that there is no intentional heating or cooling of the suspended island electrons. This guarantees that the island temperature follows the bath except for noise heating, which can be seen as the saturation of the island electron temperature at low bath temperatures. From theory \cite{giazotto06} it is expected that at low temperatures the relation between the probe voltage and the electron temperature is linear, which allows us to extrapolate the calibration to lowest temperatures. The island temperature as a function of the bath temperature at zero cooler bias voltage is shown in Fig.~\ref{fig2}(b).

\begin{figure}
\begin{center}
\includegraphics[width=1\linewidth]{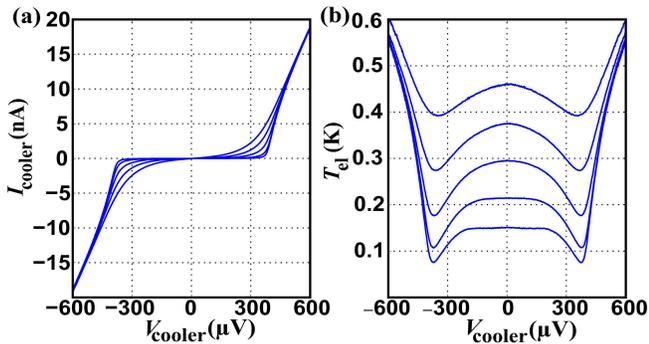}
\caption{(\textbf{a}) Measured current-voltage curves of the cooler junctions at different bath temperatures.
(\textbf{b}) Suspended island electron temperatures as function of the cooler bias voltage at same temperatures as in (a). The cooling curves and IV curves are measured simultaneously using the configuration shown in Fig.~\ref{fig1}. The bath temperatures are (from bottom to top) 103 mK,  211  mK,  295 mK,  373  mK and 460 mK. As the bath temperature is lowered, the IV curves get sharper.}
\label{fig3}
\end{center}
\end{figure}

Figure~\ref{fig3}(a) illustrates some of the measured current-voltage characteristics of the cooler junctions. The BCS gap is extracted by fitting the theoretical I-V curves to the measurement data, taking into account the charging energy \cite{saira}, measured to be roughly 190 mK. The gap was determined to be $\Delta/e\approx 194\pm 6$ $\mu$V. This is in accordance with $\Delta$ normally measured in aluminum films of such thickness. The dependence of $T_{\rm el}$ on the cooler voltage $V_{\rm cooler}$ at different bath temperatures is shown in Fig.~\ref{fig3}(b). The cooler functions well: at best the cooled electron temperature ($V_{\rm cooler} \approx 400\mu$V) is a factor of 2 lower than the temperature at zero bias ($V_{\rm cooler}=0$). From the lowest zero bias temperature of 145 mK the electron gas cools to roughly 80 mK.

In bulk samples at sub-kelvin temperatures the electrons are effectively decoupled from the ambient phonons, and the power flowing from electrons to phonons is given by \cite{wellstood} $P=\Sigma\mathcal{V}(T_{\rm el}^5-T_{\rm ph}^5)$, where $\mathcal{V}$ is the volume of the metal, $\Sigma$ is a material parameter and $T_{\rm el}$ and $T_{\rm ph}$ are the electron temperature and phonon temperature, respectively. However, at the lowest temperatures our sample's transverse dimensions become smaller than the thermal wavelength of the phonons, defined as $(hc_l)/(k_BT)$ where $c_l$ is the speed of sound, hence, affecting the phonon dimensionality and presumably this power law \cite{geller05, IlariMembrane, hekking}. In our geometry, this criterion means that at temperatures below roughly 450 mK the phonon system should be fully one dimensional, assuming that the speed of sound in AuPd is close to that of Au and Pd.

In the present experiments we see no clear evidence of modified electron-phonon coupling but the data are still inconclusive. In Fig.~\ref{fig2}(b) the saturation of the electron temperature due to noise heating is illustrated. The solid line is a fit assuming the conventional bulk electron-phonon coupling and that the island phonon temperature follows the bath temperature. The dashed line in the inset shows similar fit with $T^3$ coupling, which is expected for purely one dimensional phonons \cite{hekking}. The conventional coupling gives a better fit and using literature value of $\Sigma=2\times10^9$ ${\rm WK}^{-5}{\rm m}^{-3}$ and volume of $\mathcal{V}=10^{-20}$ m$^3$ we get a heating power of 1 fW which is in accordance to our previous experiences with the measurement apparatus. We also calculated the theoretical cooling/heating power for the whole bias range with the parameters extracted from the fits to the I-V curves. However, it turned out that the input impedance of our current amplifier is comparable with the dynamical resistance of our sample at the gap edges. This resistance covers four orders of magnitude and makes a reliable analysis of the data over the whole bias range very difficult. At low bias voltages, i.e., at the high-impedance regime, we see hints of a $T^3$ power law but the data are still inconclusive as we cannot extend this fit to the heating regime. %Also, by doing a linear fit to the $T_{\rm el}^5 - T_{\rm bath}^5$ data for the whole bias range, we can get $\Sigma=3\times10^9$ ${\rm WK}^{-5}{\rm m}^{-3}$ in reasonable agreement with the known bulk value, but scatter in the plot (not shown) is large. We also fabricated smaller SINIS devices with a Pd island using a process similar to Ref.~\cite{LiSET} and found no clear difference between suspended and unsuspended samples.

In conclusion, we have fabricated a suspended SINIS cooler and demonstrated electronic cooling of a nanoscale metallic beam. For this purpose we developed a simple single lithography step process for fabricating released metallic nanostructures with tunnel junctions. In future these devices might be of interest as bolometric detectors or in the studies of the quantum limit of nanomechanics. Also the effects of reduced phonon dimensionality in these structures are a topic of considerable interest for future experiments. This work is supported by the Academy of Finland and the NanoSciERA project "NanoFridge" of the EU. JST acknowledges partial support of Japan Science and Technology Agency through the CREST Project. We thank H. Im and T. F. Li for their assistance in fabrication of the devices.

\end{document}